\documentclass[final,5p,times,twocolumn]{elsarticle}

\usepackage{amssymb,amsmath}
\usepackage{textcomp}
\usepackage{subfigure}
\usepackage{braket}
\usepackage{float}
\usepackage{xcolor}
\usepackage{cleveref}
\usepackage{multirow}
\usepackage{icomma}
\usepackage{stackrel}

\usepackage{cleveref}

\newcounter{thmcounter}
\newtheorem{thm}{Theorem}%[thmcounter]

\newcounter{lemcounter}
\newtheorem{lem}{Lemma}%[lemcounter]

\newcounter{propcounter}
\newcounter{corcounter}
\newtheorem{cor}{Corollary}[thm]

\newcounter{defcounter}
\newtheorem{dfn}{Definition}%[defcounter]

\Crefname{thm}{Theorem}{Theorem}
\Crefname{lem}{Lemma}{Lemm}
\Crefname{prop}{Axiom}{Axiom}
\Crefname{cor}{Corollary}{Corollary}
\Crefname{dfn}{Definition}{Definition}

\begin{document}

\title{Fractal derivatives, fractional derivatives and $q$-deformed calculus}

\author{Airton Deppman}
 \address{Instituto de Física -  Universidade de São Paulo, São Paulo 05508-090, Brazil.%\url{https://orcid.org/0000-0001-9179-6363}
 }
\ead{deppman@usp.br}

\author{Eugenio Meg\'{\i}as}
 \address{Departamento de Física Atómica, Molecular y Nuclear and Instituto Carlos I de Física Teórica y Computacional, \\
 Universidad de Granada, Avenida de Fuente Nueva s/n, 18071 Granada, Spain
 }
\ead{emegias@ugr.es}

\author{Roman~Pasechnik}
\address{Department of Physics, Lund University, S\"olvegatan 14A,
Lund SE-22362, Sweden}
\ead{Roman.Pasechnik@hep.lu.se}

\begin{abstract}
This work presents an analysis of fractional derivatives and fractal derivatives, discussing their differences and similarities. The fractal derivative is closely connected to Haussdorff's concepts of fractional dimension geometry. The paper distinguishes between the derivative of a function on a fractal domain and the derivative of a fractal function, where the image is a fractal space. Different continuous approximations for the fractal derivative are discussed, and it is shown that the $q$-calculus derivative is a continuous approximation of the fractal derivative of a fractal function. A similar version can be obtained for the derivative of a function on a fractal space. Caputo's derivative is also proportional to a continuous approximation of the fractal derivative, and the corresponding approximation of the derivative of a fractional function leads to a Caputo-like derivative. This work has implications for studies of fractional differential equations, anomalous diffusion, information and epidemic spread in fractal systems, and fractal geometry.
\end{abstract}

\maketitle

\section{Introduction}
Fractional differential equations have been used to describe the behavior of complex systems. The~growing interest in this mathematical tool imposes the necessity of urgent analysis of its fundamentals. The~widespread use of fractional differential equations in fluid dynamics, finance, and~other complex systems has led to the intense investigation of the properties of fractional derivatives and their geometrical and physical meaning. Fractional derivatives are often associated with fractal geometry, but~the connections between fractional derivatives and fractal derivatives have not been clarified so far. Fractional derivatives have been used in many applications~\cite{Tenreiro_Machado2010-up,Debnath2003-fc}, and~advancing our understanding of their geometrical meaning and their relations with fractals is necessary. The~$q$-calculus has been frequently applied to describe the statistical properties of fractal systems~\cite{Deppman2021-ya,Deppman2021-vt}. However, the~relationship between $q$-calculus and fractal derivatives has not been fully understood~yet.

This work reviews the fundamentals of fractal derivatives and establishes their connections with fractional derivatives and $q$-calculus. The~generalization of standard calculus to include fractional-order derivatives and integrals is an exciting field of research, and~many works have been conducted in this area. Different proposals for fractional generalization are available, and~applications of fractional derivatives have been used in various fields. Fractional differential equations are frequently used to describe the behavior of complex systems. In~Refs.~\cite{Valerio2013-mu,Valerio2022-tt}, the~authors analyzed different forms of fractional derivatives and discussed their properties. Caputo's derivative is among the most commonly used and is defined by
\begin{equation}
  D_C^{\nu} h(x)=\frac{1}{\Gamma(1-\nu)}\int_{x-\delta}^{x} (x-t)^{-\nu} \frac{dh}{dt} dt \,,
 \end{equation}
 which is a particular case of the Riemann--Liouville fractional derivative~\cite{Abdeljawad2011-lt}. 

Haussdorff established the fundamental aspects of spaces with fractional dimension, and~an introduction to the subject can be found in~\cite{Falconer2014-de}. One of the important quantities associated with fractal spaces is the Haussdorff measure, denoted by ${\cal H}^{s}(\mathbb{F})$. Its definition is based on the measure ${\cal H}^{s}_{\delta}(\mathbb{F})$, and~is given by
\begin{equation}
{\cal H}^{s}(\mathbb{F})=\lim_{\delta \rightarrow 0}{\cal H}^{s}_{\delta}(\mathbb{F}) \,,
\end{equation}
where the measure depends on a $\delta$-cover of the Borel subset $\mathbb{F} \subseteq \mathbb{R}^n$. The~space $\mathbb{F}$ will be referred to as a fractal space, and~its Hausdorff dimension is denoted by $\alpha$ and defined as
\begin{equation}
\alpha= \inf\{s \ge 0:{\cal H}^{s}(\mathbb{F})=0 \}= \sup\{s: {\cal H}^{s}(\mathbb{F})=\infty \}\,.
\end{equation}

If $0<\alpha<\infty$, the~Haussdorff measure of the $\delta^\alpha$-cover is called the mass distribution, denoted by $\gamma^\alpha(\mathbb{F},a,b)$~\cite{ParvateGangal1,ParvateGangal2,ParvateSatinGangal}, which will be discussed below. Fractal derivatives and fractional derivatives are not the same concept~\cite{Golmankhaneh2016-oh}, and~the non-locality is a prominent aspect of the fractal derivative. For~a comprehensive review of the subject and its applications, see Ref.~\cite{AlirezaBook}. The~Parvate--Gangal derivative is defined for functions on a fractal domain. This work shows that extending the same concepts to functions with a fractal image can provide new insights into the role of fractal derivatives in the study of complex~systems.

Tsallis statistics was proposed to describe the statistical properties of fractal systems. It introduces a non-additive entropy that can be used to obtain, through the ordinary thermodynamics formalism, the~non-extensive thermodynamics~\cite{Tsallis,TsallisBook}. To~deal with non-additivity, the~$q$-calculus was proposed~\cite{BORGES-qCalculus}. One important result of $q$-calculus is the $q$-derivative, which is written as:
\begin{equation}
\frac{\bar{d} f}{dx}=f^{q-1}\frac{df}{dx}. \label{q-derivative}
\end{equation}
Notice that, if~the function $f$ is a $q$-exponential, the~{\it special derivative} above results to be identical to the standard derivative of a $q$-exponential function. This derivative can be straightforwardly related to the conformal derivative~\cite{Khalil2014-fo}.

The three different theoretical areas mentioned above have been investigated independently, evolving in parallel. Despite their many common aspects, the~connections between them have not been demonstrated so far~\cite{Cattani2017-ul}. This work aims to establish connections between Caputo's derivative and the $q$-calculus with the continuous approximation of the fractal derivative proposed by Parvate and Gangal. {In this %MDPI: The bold format has been removed, please confirm. R: confirm.
 work, we assume that the fractal derivative is correctly calculated by the definitions advanced by Parvate, Gangal, and coworkers~\cite{ParvateGangal1,ParvateGangal2,ParvateSatinGangal}, and~discuss how some relevant forms of fractional derivatives, as~well as the $q$-deformed derivative, can be obtained as a continuous approximation of the fractal~derivative.}

\section{Fractal~Derivatives}

\vspace{6pt}
\begin{lem}
 If $x=(x_1, \cdots , x_n) \in \mathbb{R}^n$ and $f=(f_1(x), \cdots , f_m(x)) \in \mathbb{R}^m$ is an $m$-dimensional vector field $f: \mathbb{R}^n \rightarrow \mathbb{R}^m$. Then, $m \le n$.
 \label{lem:DomainImageDimension} 
\end{lem}

{\bf Proof:} Suppose $m > n$, then $(f_1(x), \cdots, f_n(x))$ forms a new set of $n$ independent
variables, which are functions of the $n$ independent variables
of $x$. Then, $f_{n+1}(x)$ is not independent of the functions in the~set.
 
 \vspace{0.5cm}

\begin{dfn}
 A vector field with dimension $m=1$ is a function.
\end{dfn}

\begin{lem}
 If there is an inverse function $f^{-1}\left(f(x)\right)=x$, then $m=n$.
\end{lem}

{\bf Proof:} It follows immediately by applying \Cref{lem:DomainImageDimension}.

\begin{lem}
 If $f$ is a fractal vector field $f:\mathbb{R}^n \rightarrow \mathbb{R}^{\alpha}$, with~$\alpha \in \mathbb{R}$, then $\alpha \le n$.
\end{lem}
 
 {\bf Proof:} It follows immediately by applying \Cref{lem:DomainImageDimension}.

\begin{dfn}
 A fractal vector field with dimension $\alpha \le 1$ is a fractal function.
\end{dfn}

\begin{dfn}
 An $\alpha$-dimensional function is a fractal vector field if $\alpha > 1$ or a fractal function if $\alpha \le 1$.
\end{dfn}

\begin{dfn} If $\gamma(\mathbb{F},~a,~b)$ is the Haussdorff mass distribution for a cover $F$, with~$a, b \in F$, then the staircase function, $S^{\alpha}_{F,a_o}$, is defined as
\begin{equation}
  S^{\alpha}_{F,a_o}=
    \begin{cases}
        \gamma(\mathbb{F},a_o,x) ~~\textrm{for $x > a_o$} \\
        \gamma(\mathbb{F},x,a_o) ~~\textrm{for $x < a_o$}
    \end{cases}\,.
 \end{equation} \label{dfn:staircase}
\end{dfn}

\vspace{-23pt}
\begin{lem}
    The staircase function is a scalar.
\end{lem}

{\bf Proof:} The staircase function is proportional to the Haussdorff mass function, which is the volume resulting from the union of the $\delta^{\alpha}(x) \in \mathbb{F}$, so it is a~scalar.

\begin{dfn} If $\mathbb{F}$ is a $\delta^{\alpha}$-cover and $f: \mathbb{F} \rightarrow \mathbb{R}$, then the fractal derivative of $f(x)$ is
\begin{equation}
    D_{\mathbb{F},a_o}^{\alpha}f(x_o)=
    \begin{cases}
     F\lim_{x \rightarrow x_o}\frac{f(x)-f(x_o)}{S_{F,a_o}^{\alpha}(x)-S_{F,a_o}^{\alpha}(x_o)} ~~~~x, x_o\in \mathbb{F} \\
    0 ~~~~\textrm{otherwise}
    \end{cases}\,.
\end{equation} \label{dfn:Dalpha}
\end{dfn}

\vspace{-26pt}
\begin{thm}
 There is a fractal derivative of the inverse function, and~it is the inverse of the fractal~derivative.
\label{thm:invD}
\end{thm}

 {\bf Proof: } Consider that $x, x_o \in \mathbb{F}$. Suppose there exists a function $g: \mathbb{R} \rightarrow  \mathbb{F} $ such that $g(f(x))=x$. Then,
\begin{equation}
D_{\mathbb{F},a_o}^{\alpha}g\left(f_{x_o}\right)=F\lim_{x \rightarrow x_o}  \frac{g(f_x)-g(f_{x_o})}{f(x)-f(x_o)}   \frac{f(x)-f(x_o)}{S_{F',a_o}^{\alpha}(x)-S_{F',a_o}^{\alpha}(x_o)} =1\,,
\end{equation}
where the simplified notation $f_x=f(x)$ was adopted. It follows that
\begin{equation}
 F\lim_{x \rightarrow x_o}  \frac{g(f_x)-g(f_{x_o})}{f(x)-f(x_o)}=
 F\lim_{x \rightarrow x_o}\frac{S_{F', a_o}^{\alpha}(x)-S_{F', a_o}^{\alpha}(x_o)}{f(x)-f(x_o)} \,.  \label{eq:F1}
\end{equation} 

The fractal derivative of the inverse function can be applied to any fractal function $h$: $\mathbb{R} \rightarrow \mathbb{F}$. The~staircase function, in~this case, is applied to the fractal image space of the function~$h$. The~function $f$ can be defined arbitrarily, with~the constraint that there is an inverse function $f^{-1}$. One case of particular interest is the identity function $f(x)=x$, then we have
\begin{equation}
[D_{\mathbb{F},\varphi}^{\alpha}]^{-1}h(x_o)=F\lim_{x \rightarrow x_o}\frac{S_{F, \varphi}^{\alpha}[h(x)]-S_{F, \varphi}^{\alpha}[h(x_o)]}{x-x_o}\,,
\end{equation}
with $\varphi=h(a_o)$.

Observe that in this case, the~image space and the domain space of the function $h$ are the same, i.e.,~$h$: $\mathbb{F} \rightarrow \mathbb{F}$.

\begin{dfn}
The result obtained above can be  generalized by defining the fractal derivative of the inverse function or, equivalently, the~inverse of the fractal derivative, as~\begin{equation}
[D_{\mathbb{F},\varphi}^\alpha]^{-1}h(f_{x_o}) = 
\begin{cases}
  F\lim_{x \rightarrow x_o}\frac{S_{F,\varphi}^{\alpha}[h(x)]-S_{F,\varphi}^{\alpha}[h(x_o)]}{x-x_o} ~~~~x, x_o\in \mathbb{F}\,. \\
  0 \qquad \textrm{ otherwise}
\end{cases}
 %\,. 
\end{equation}
\end{dfn}

\begin{cor}
 The derivative of a fractal function is well-defined only if the function is almost always non-divergent in the interval $[a, b]$~(Following the standard terminology in the field, we say that a function is almost always non-divergent if the set of points where it is divergent has null Lebesgue measure).
\end{cor}

{\bf Proof:} According to~\Cref{dfn:staircase}, the~staircase function is well-defined only if the mass distribution function can be defined. The~mass distribution is equal to the Haussdorff measure when the Haussdorff dimension is $0<\alpha< \infty$. This condition is satisfied only if the function is almost always~non-divergent. 

\vspace{0.5cm}

\begin{thm} \label{thm:DerDim}
If the function $h(x)$ is almost always continuous and non-divergent in $\mathbb{F}$, and $h'(x)=[D_{\mathbb{F},\varphi}^\alpha]^{-1}h(x)$, then the Haussdorff dimension of $h(x)$ and $h'(x)$ are the same.
\end{thm}

{\bf Proof:} Let $\mathbb{F}$ be the $\delta^{\alpha}$-cover of the fractal function $h(x)$, and~$\mathbb{F}'$ the $\delta^{\beta}$-cover of the inverse of fractal derivative. For~any $\delta^{\alpha}[h(x)] \in \mathbb{F}$ there is a $\delta^{\beta}[h'(x)] \in \mathbb{F}'$, so $\beta \ge \alpha$. For~$\delta^{\beta}[h'(x)] \in \mathbb{F}'$, there is a $\delta^{\alpha}[h(x)] \in \mathbb{F}$; therefore, $\alpha \le \beta$. Hence,  $\alpha=\beta$.

\vspace{0.5cm}

\begin{dfn}
 We will denote the inverse of an $\alpha$-dimensional fractal function by $D_{\mathbb{F},\varphi}^\alpha h(x)$, and~we will refer to it as a fractal derivative of an $\alpha$-dimensional fractal function, or~simply fractal function, while the fractal derivative will be called fractal derivative over a fractal space.
\end{dfn}

\vspace{0.5cm}

\begin{dfn}
 The partial derivative of a fractal function is
\begin{equation}
  D_{\mathbb{F},\varphi}^\alpha|_ih(f_{x})=
  \begin{cases}
  F\lim_{x_i \rightarrow x_{o,i}}\frac{S_{F,\varphi}^{\alpha}[h(x)]-S_{F,\varphi}^{\alpha}[h(x_o)]}{x_i-x_{o,i}} ~~~~x, x_o\in \mathbb{F}  \,,\\
  0 \qquad \textrm{ otherwise}
\end{cases}
 \end{equation}
where the index $i$ indicates the component $x_i$ %MDPI: We changed the ``$x^i$'' into ``$x_i$'', please confirm. R: I confirm.
 of the vector $x$. 
\end{dfn}

\vspace{0.5cm}

\begin{cor}
 The dimension of $D_{\mathbb{F},\varphi}^\alpha|_i \, h(f_{x})$ is $\alpha \le 1$.
\end{cor}

{\bf Proof:} It follows immediately from \Cref{lem:DomainImageDimension} and \Cref{thm:DerDim}.

\vspace{0.5cm}

\begin{dfn}
 The staircase function differential is defined by
\begin{equation}
   dS_{F,a_o}^{\alpha}(x)=
   \begin{cases}
     F \lim_{dx \rightarrow 0} \left[S_{F,a_o}^{\alpha}(x+dx)-S_{F,a_o}^{\alpha}(x)\right] \textrm{ if $x, x+dx \in \mathbb{F}$} \\
     0 \qquad \textrm{ otherwise}
   \end{cases}
 \end{equation}
\end{dfn}

\vspace{0.5cm}

\begin{thm} The staircase function differential can be approximated by
\begin{equation}
    dS_{F,a_o}^{\alpha}(x)=\frac{A(\alpha)}{\alpha}dx^{\alpha}\,,
\end{equation} 
where
\begin{equation}
 A(\alpha) := 2\pi^{\alpha/2}/\Gamma(\alpha/2)\,. \label{eq:Aalpha}
\end{equation}
\end{thm}

{\bf Proof:} For any volume $(\delta x)^n \in \mathbb{R}^n$, its intersection with $\mathbb{F}$ has a volume $(\delta x)^{\alpha}$. Consider the volume of an $n$-dimensional sphere of radius $x$ given by
\begin{equation}
 V(x) = \frac{A(n)}{n} x^n \,,
\end{equation}
where $A(n) = 2\pi^{n/2}/\Gamma(n/2)$ is the surface area term, with~$\Gamma(z)$ being the Euler's Gamma Function, and~$x = \sqrt{x_1^2+ \cdots + x_n^2}$.
Then, the~volume of a spherical shell of finite width $\delta x$ is given by
\begin{equation}
 \delta V(x) = \frac{A(n)}{n} \left( (x + \delta x)^n - x^n \right) \,.
\end{equation}
\noindent{In the limit $\delta x \rightarrow dx$, where now $dx$ is infinitesimal, it results}
\begin{equation}
 dV(x)= A(n) x^{n-1} dx = \frac{A(n)}{n} dx^n \,,
\end{equation}
where $dx^n := d(x^n)$.

The intersection of $\delta V(x)$ with $\mathbb{F}$, which is denoted by $\delta V_{\alpha}(x)$, is
\begin{equation}
 \delta V_{\alpha}(x)=\frac{A(\alpha)}{\alpha} \left( (x + \delta x)^{\alpha} - x^\alpha \right) \,.
\end{equation}
In the limit $\delta x \rightarrow dx$, this leads to
\begin{equation}
\delta V_{\alpha}(x) \rightarrow d V_{\alpha}(x) = A(\alpha) x^{\alpha - 1} dx = \frac{A(\alpha)}{\alpha} d x^{\alpha} \,.
\end{equation}
On the other hand, $dS_{F,a_o}^{\alpha}(x)$ is the volume of the intersection between an infinitesimal volume $dV \in \mathbb{R}^n$ with $\mathbb{F}$. (The multiplicative coefficient $A(\alpha)$ used here is valid for integer dimensions. The~case of fractional dimensions is more challenging, so this coefficient needs to be considered with care. In~this work, we focus on the shape of the continuous approximation.) 

\begin{equation}
    dS_{F,a_o}^{\alpha}(x)=\frac{A(\alpha)}{\alpha}dx^{\alpha}=A(\alpha)x^{\alpha-1} dx\,. \label{dS}
\end{equation}

\vspace{0.5cm}

\begin{dfn}
 The continuous approximation of a fractal function is defined as a set of infinitesimal elements $dx$ such that \Cref{dS} is satisfied. \label{dfn:CP}
\end{dfn}

\vspace{0.5cm}

\begin{thm}
    The continuous approximation of the fractal derivative of a function is
\begin{equation}
    D^{\alpha}_{\mathbb{F},\varphi} h(x)=\frac{A(\alpha)}{\alpha}\frac{dh^{\alpha}}{dx}=A(\alpha)h^{\alpha-1}(x)\frac{dh}{dx}(x) \,.
 \end{equation} \label{thm:CPd}
\end{thm}

\vspace{0.5cm}

\begin{thm}
    The continuous approximation of the fractal derivative of a fractal function is
\begin{equation}
        D^{\alpha}_{\mathbb{F},\varphi} h(x)=\frac{A(\alpha)}{\alpha}\frac{dh^{\alpha}}{dx}=A(\alpha)h^{\alpha-1}(x)\frac{dh}{dx}(x) \,.
    \end{equation} \label{thm:CPf}
\end{thm}

{\bf Proof:} It follows directly from the definition of the fractal derivative of a function and of the continuous approximation.

\vspace{0.5cm}

\begin{thm}
 Consider a fractal function $f: \mathbb{R}^n \rightarrow \mathbb{F}$, where $\mathbb{F}$ is a $\delta^{\alpha}$-cover, with~$n-1<\alpha<n$, for~$n>1$. It defines a set of fractal functions $\{f_i(x_i)\}$ with dimensions $\{\alpha_i\}$ such that \mbox{$\alpha=\alpha_1+ \cdots + \alpha_n$}.
\end{thm}

{\bf Proof:} Consider the fractal function $f_k(x_k)=f(a, \cdots, x_k, \cdots, z)$, where $a , \cdots , z$ are constants. For~any interval $I=[x_k, x_k+\delta x_k]$, the~intersection of $I$ and $\mathbb{F}$ is $(\delta x_k)^{\alpha_k}$, with~$\alpha_k < 1$. For~an $\alpha_{k-1}$-dimensional function $h_{k-1}(x_1, \cdots , x_{k-1}, k, l, \cdots , z)$ such that for any volume $(\delta x)^{k-1}$, the~intersection with $\mathbb{F}$ is $(\delta x)^{\alpha_{k-1}}$, the~function $h_k(x_1, \cdots , x_{k-1}, x_k, l, \cdots , z)$ has dimension $(\delta x)^{\alpha_{k-1}} \delta x=(\delta x)^{\alpha_{k}}$, where $\alpha_k=\alpha_{k-1}+\alpha_k$. The~theorem is proved by induction.

\vspace{0.5cm}

\begin{dfn}
 Consider a fractal function $h$ with dimension $\alpha < 1$. The~gradient of a fractal function is defined as
\begin{equation}
  {\bf D}_{\mathbb{F},\varphi}^{\alpha}h(x_o)=\left(D_{\mathbb{F},\varphi}^{\alpha_1}|_1 \, h(x_o), \cdots , D_{\mathbb{F},\varphi}^{\alpha_n}|_n \, h(x_o) \right) \,,
 \end{equation}
 where $\alpha=\alpha_1 + \cdots + \alpha_n$.
\end{dfn}

\vspace{0.5cm}

\begin{dfn}
 For $\alpha>1$, the~partial fractal derivative of the function is
\begin{equation}
  {\bf D}_{\mathbb{F},\varphi}^{\alpha}|_i h(x_o)=\left(D_{\mathbb{F},\varphi}^{\alpha_1}|_i \, h(x_o), \cdots , D_{\mathbb{F},\varphi}^{\alpha_n}|_i \, h(x_o) \right) \,,
 \end{equation}
 where $\alpha=\alpha_1 + \cdots + \alpha_n$.
\end{dfn}

\vspace{0.5cm}

\begin{thm}
 For a finite $\delta$, the~derivative of a fractal function in the interval $[x-\delta, x]$ is
\begin{equation}
   D_{[\delta],\varphi}^{\alpha} h(x)=\frac{A(\alpha)}{\alpha}\int_{x-\delta}^{x} h^{\alpha-1}(t)\frac{dh}{dt} dt \,.   \label{eq:Dhphi}
 \end{equation}
\end{thm}

{\bf Proof:} The derivative in the interval $[x-\delta, x]$ is
\begin{equation}
 D_{[\delta],\varphi}^{\alpha} h(x)= \int_{x-\delta}^{x} D_{\mathbb{F},\varphi}^{\alpha} h(t) dt \,.
\end{equation}
Using \Cref{dfn:CP}, the~theorem is proved.

\vspace{0.5cm}

\begin{thm}
For a finite $\delta$, the~derivative of function in the interval $[x-\delta, x]$ in a fractal space is
\begin{equation}
   D_{[\delta],a}^{\alpha} h(x)=\frac{A(\alpha)}{\alpha}\int_{x-\delta}^{x} [h(x)-h(t)]^{\alpha-1}\frac{dh}{dt} dt  \,. 
 \label{eq:Dha}
 \end{equation}
\end{thm}
 
{\bf Proof:} The proof is performed by applying the continuous approximation in \Cref{dS} to the derivative on fractal space in \Cref{dfn:Dalpha}.

\vspace{0.5cm}

Observe that the $\alpha$-dimensional sphere needs not to be centered at $\varphi$ for the fractal derivative of a fractal function, or~at $a$ for the derivative on a fractal space. The~point $x$, where the derivative is calculated, can be set as the center of the~sphere.

\vspace{0.5cm}

\vspace{0.5cm}

\begin{dfn} 
 The continuous approximation of the derivative of a function on a fractal space, based on $\alpha$-dimensional sphere centered at $x$ is indicated by $D_{\mathbb{F},x}^{\alpha} h(x)$. \label{dfn:NL-CP}
\end{dfn}

\vspace{0.5cm}

\begin{thm} \label{thm:CPsnl}
 The continuous approximation of the derivative of a function on a fractal space, $D_{\mathbb{F},x}^{\alpha} h(x)$ in the interval $[x-\delta,x]$, for~finite $\delta$, is given by
\begin{equation}
  D_{\mathbb{F},x}^{\alpha} h(x)=\frac{A(\alpha)}{\alpha}\int_{x-\delta}^{x} (x-t)^{1-\alpha} \frac{dh}{dt} dt \,, 
 \end{equation}
 which is proportional to Caputo's derivative. \label{thm:Caputo} 
\end{thm}

{\bf Proof:} The local continuous approximation, considering that the radius of the spherical shell is $x-t$, is determined from \Cref{thm:CPf} as
\begin{equation}
    D^{\alpha}_{\mathbb{F},x} h(t)=A(\alpha)(x-t)^{1-\alpha}\frac{dh}{dx}(t)\,.
\end{equation}

\noindent{Using \Cref{dfn:NL-CP}, one has}
\begin{equation}
     D^{\alpha}_{\mathbb{F}}h(x) =  \int_{x-\delta}^x D^{\alpha}_{\mathbb{F},x} h(t) dt \,, \label{eq:proof_caputo}
\end{equation}
leading to the proof of the~Theorem. 

 \vspace{0.5cm}

\begin{dfn}
 The continuous approximation of the derivative of a fractal function based on $\alpha$-dimensional sphere centered at $x$ is indicated by $D_{\mathbb{F}}^{\alpha} h(x)$. \label{dfn:NL-CPf}
\end{dfn}

\begin{thm} \label{thm:Caputo-like}
 The continuous approximation of the derivative of a fractal function, $D_{\mathbb{F},\varphi_x}^{\alpha} h(x)$ in the interval $[x-\delta, x]$, for~finite  $\delta$, is given by
\begin{equation}
    D_{\mathbb{F},\varphi_x}^{\alpha} h(x)=\frac{A(\alpha)}{\alpha}\int_{x-\delta}^{x} (\varphi_x-h(t))^{\alpha-1} \frac{dh}{dt} dt \,, \label{eq:DFa}
 \end{equation}
\end{thm}
for $t$ such that $h(t)<\varphi_x=h(x)$.

{\bf Proof:} The proof follows the same lines of the proof  {for \Cref{thm:Caputo}.

\begin{cor}
 The continuous approximation in \Cref{dfn:CP} is proportional to the limit of the continuous approximation in the range $[x-\delta, x]$ for $\delta \rightarrow 0$ of Caputo's derivative.
\end{cor}

\section{Discussion and Conclusions}

The fractal derivative proposed by Parvate and Gangal, presented in \Cref{dfn:Dalpha}, is the closest concept to the Hausdorff concept of fractional dimension spaces. Therefore, it is considered as the starting point for the analysis of fractal derivatives and fractional derivatives~here.

The existence of the inverse of the Parvate--Gangal derivative is a natural consequence, i.e.,~a derivative of a function with a fractal image space that is defined on a domain space, which may or may not be fractal. This is proven in \Cref{thm:invD}.

This work demonstrates that fractal functions with arbitrary dimension $\alpha$, such as a fractal vector field with fractal dimension $\alpha>1$, can be defined. However, the~cases of most interest are those with $\alpha \le 1$, as~they are physically relevant for the present~work.

The derivative of a fractal function on a fractal space allows for a continuous approximation, as~demonstrated in \Cref{thm:CPd}. Additionally, a~similar continuous approximation can be obtained for the derivative of a function in a fractal space, as~shown in \Cref{thm:CPf}. This approximation is identical to the {\it special derivative} used in Ref.~\cite{Deppman:2023ztz} to derive the Plastino--Plastino Equation, which is a generalization of the Fokker--Planck Equation for systems with non-local~correlations.

To illustrate the behavior of the continuous approximation, we utilize the well-known Cantor Set, which has a dimension $\alpha=0.631$.  We aim to demonstrate how the continuous approximation aligns with the mass distribution, $S^{\alpha}_{\mathbb{F},0}(x)$. To~achieve this, we numerically calculate the mass distribution for this fractal set up to level 4. In~other words, the~smallest component of the fractal has a linear length of $l=3^{-4}$. We employ a $\delta$-cover with $\delta=0.01$ to calculate the mass~distribution. 

Next, we fit a power-law function, $y(x)=ax^b$. According to the theoretical findings presented in this work, the~exponent $b$ should closely approximate the fractal dimension $\alpha$ of the Cantor Set. The~obtained results are displayed in Figure \ref{fig:CS-CA}, revealing that the best fit corresponds to $b=0.636$, which is in close proximity to the expected value. This outcome effectively illustrates the application of the continuous approximation and provides insight into substituting the mass distribution by the continuous approximation. It should be noted that there are numerous other methods available for creating a continuous approximation of the fractal measure, and~each of these approaches will result in different fractional derivatives. Investigating the coherence and convenience of different forms of approximation to the staircase function is an interesting line of research that is beyond the scope of this present study.

%\vspace{-11pt}

\begin{figure}[t]
%\centering
\includegraphics[width=0.5\textwidth]{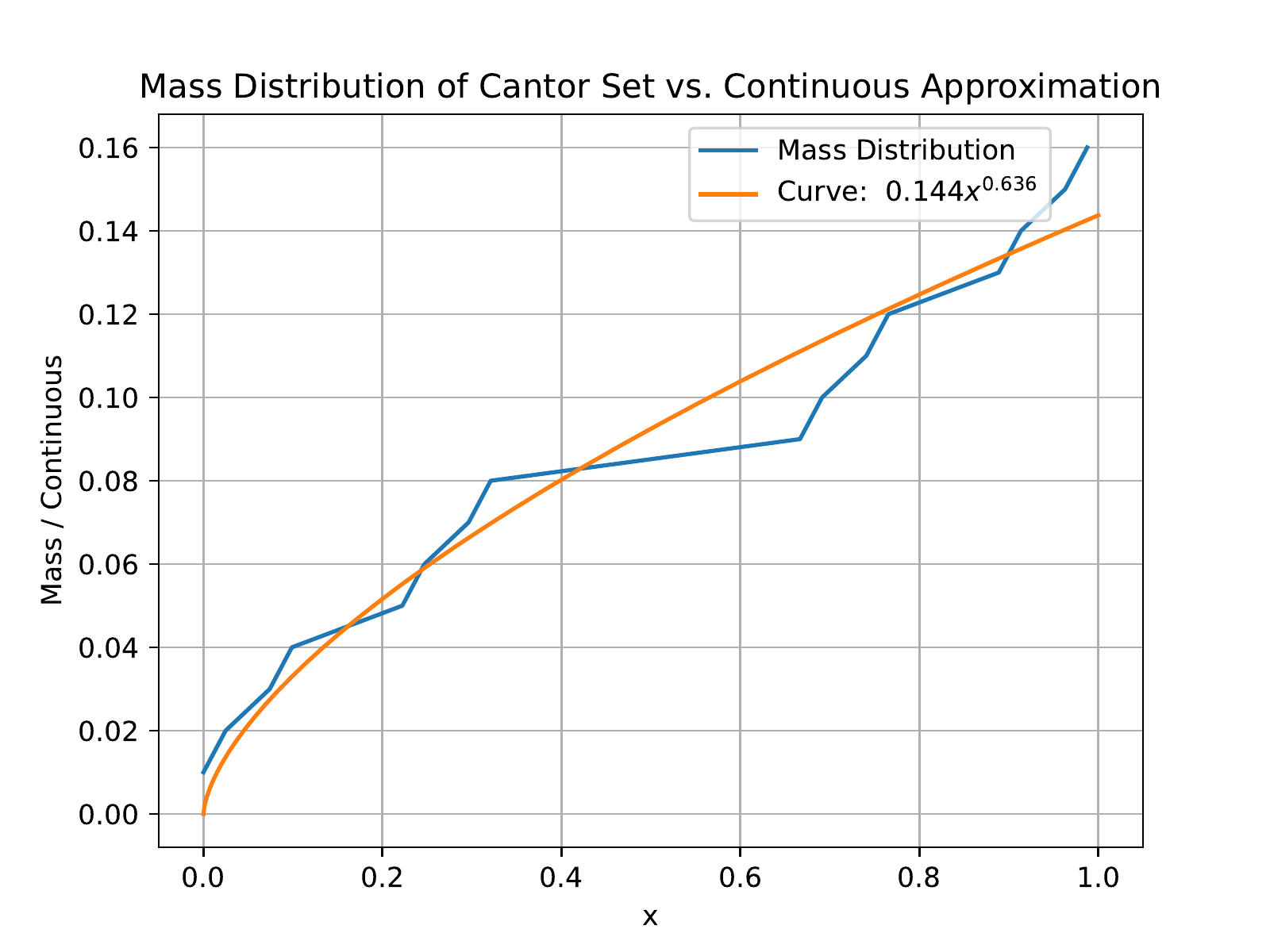}  
 \caption{Plots of the mass distribution (blue line) for the Cantor Set at the 4th iteration, calculated with a $\delta$-cover with $\delta=0.01$, compared with the continuous approximation (orange line) represented by a function $y(x)=ax^b$ fitted to the mass distribution. The~best-fit results in $b=0.636$, in~agreement with the Cantor Set dimension $\alpha=0.631$.}
\label{fig:CS-CA}
\end{figure}

The continuous approximation derivative is expressed in terms of the standard derivative operator and can be associated with the $q$-deformed calculus~\cite{BORGES-qCalculus}. Unlike the fractal derivative, the~continuous approximation is a local derivative, and~the non-linear behavior of the continuous approximation is a remnant of the non-local properties of the fractal~derivative.

Non-locality can be explicitly introduced into the continuous approximation by considering finite $\delta$-covers. In~the non-local continuous approximation, the~derivative is obtained by integrating the local continuous derivative over a finite range $\delta$. This non-local continuous approximation is presented in \Cref{thm:CPsnl}, and~it is precisely the Caputo fractional~derivative.

The derivative of a function in a fractal space and the derivative of a fractal function lead to different continuous approximations. The~former can be associated with the Caputo fractional derivative, as~shown in~\Cref{thm:CPsnl}, while the latter leads to a Caputo-like derivative, as~demonstrated in~\Cref{thm:Caputo-like}. Similar derivatives to Caputo's derivative can also be found in~\cite{Almeida2017-hy}.

The results of the present work evidence the relations between the fractal derivative and some of the most used fractional derivatives. 
Comparing the result of~\Cref{thm:CPf} with Equation~(\ref{q-derivative}), it is clear that the local continuous approximation of the derivative of a fractal function is equal to the $q$-derivative. Thus, for~the first time, the~$q$-calculus derivative is shown to be a continuous approximation to the fractal~derivative. 

A consequence of the relationship between the $q$-derivative and the local continuous approximation of the derivative of a fractal function (\Cref{thm:CPf}), and~of the connection between the derivative of a fractal function and the Caputo-like fractional derivative \mbox{(\Cref{thm:Caputo-like})} is that the $q$-derivative and the Caputo-like derivative are connected through a dislocation of the center of the $\alpha$-dimensional sphere around which the non-local continuous approximation is calculated. Hereby, one can conclude that different forms of fractional derivatives can be obtained from the Parvate--Gangal fractal derivative by considering the different possibilities of continuous approximation and non-locality of the fractional~derivative.

Other fractal derivatives can be explored along the same lines as performed here. The~Riemann--Liouville derivative bears a close relationship with Caputo's} derivative~\cite{Srivastava} and it is interesting to observe the similarities between the fractal derivative proposed in Refs.~\cite{He2018-yw,He2014-cd} and the continuous approximations studied in the present work. The~fractional derivative used in Ref.~\cite{Chen2006-yo} is equal to the local continuous approximation of the fractal derivative of a function in a fractal space obtained in the present work. Ref.~\cite{Weberszpil2015-ko} studied this fractional derivative and its relationship with the q-derivative. Establishing a clear connection between the Parvate--Gangal fractal derivative and Caputo's fractional derivative, this work opens the possibility for a deeper understanding of the use of fractional differential equations, which is so common in many different areas. In~this respect, let us remark that fractal and fractional differential equations have been used in applications as dynamic of the system in porous or heterogeneous media~\cite{Yin2023-ug,Bouras2023-hb,Liang2022-vf}, diffusive flow~\cite{Liu2022-oo,Wang2022-pi,Sun2023-ec,Wang2022-ch,El-Nabulsi2022-cy}, solitons~\cite{Manikandan2022-hw}, control of complex systems~\cite{Sadek2022-qq}, epidemic process~\cite{Policarpo2023-ih}, polymer plasma~\cite{Paun2022-no} and many others. The~consequences of the present study for these physical systems deserve further investigation in the future. The~consequences of the present study for these physical systems deserve further investigation in the future. Its implication on the study and applications of fractal functions~\cite{Gowrisankar2021-vi} deserves further~investigation.

\section{Acknowledgements}
The authors acknowledge fruitful discussions with Dr Alireza K. Golmankhaneh and thank M.S. Josué and M. P. Policarpo for producing the plot in Fig.~\ref{fig:CS-CA}. A.D. is supported by the Project INCT-FNA (Instituto Nacional de Ci\^encia e Tecnologia - F\'{\i}sica Nuclear Aplicada) Proc. No. 464898/2014-5, by the Conselho Nacional de Desenvolvimento Cient\'{\i}fico e Tecnol\'ogico (CNPq-Brazil), grant 304244/2018-0, by Project INCT- FNA Proc. No. 464 898/2014-5, and by FAPESP, Brazil
grant 2016/17612-7. The work of E.M. is supported by the project PID2020-114767GB-I00 funded by MCIN/AEI/10.13039/501100011033, by the FEDER/Junta de Andaluc\'{\i}a-Consejer\'{\i}a de Econom\'{\i}a y Conocimiento 2014-2020 Operational Programme under Grant A-FQM-178-UGR18, and by Junta de Andaluc\'{\i}a under Grant FQM-225. The research of E.M. is also supported by the Ram\'on y Cajal Program of the Spanish MICIN under Grant RYC-2016-20678. R.P. is supported in part by the Swedish Research Council grants, contract numbers 621-2013-4287 and 2016-05996, as well as by the European Research Council (ERC) under the European Union’s Horizon 2020 research and innovation programme (grant agreement No. 668679).

\bibliographystyle{ieeetr}
\bibliography{FracDer}

\end{document}